# A Genetic Algorithm Based Finger Selection Scheme for UWB MMSE Rake Receivers[1]

Sinan Gezici, Mung Chiang, H. Vincent Poor and Hisashi Kobayashi
Department of Electrical Engineering
Princeton University, Princeton, NJ 08544
{sgezici,chiangm,poor,hisashi}@princeton.edu

*Abstract*— Due to a large number of multipath components in a typical ultra wideband (UWB) system, selective Rake (SRake) receivers, which combine energy from a subset of multipath components, are commonly employed. In order to optimize system performance, an optimal selection of multipath components to be employed at fingers of an SRake receiver needs to be considered. In this paper, this finger selection problem is investigated for a minimum mean square error (MMSE) UWB SRake receiver. Since the optimal solution is NP hard, a genetic algorithm (GA) based iterative scheme is proposed, which can achieve near-optimal performance after a reasonable number of iterations. Simulation results are presented to compare the performance of the proposed finger selection algorithm with those of the conventional and optimal schemes.

## I. INTRODUCTION

Recently impulse radio (IR) ultra wideband (UWB) systems ([1]-[5]) have drawn considerable attention due to their suitability for short-range high-speed data transmission and precise location estimation. In an IR-UWB system, very short pulses with a low duty cycle are transmitted, and each information symbol is represented by positions or polarities of a number of pulses. Each pulse resides in an interval called "frame", and positions of pulses in frames are determined by time-hopping (TH) sequences specific to each user, which prevents catastrophic collisions among pulses of different users [1].

Commonly, Rake receivers are employed in an IR-UWB system to collect energy from different multipath components. A Rake receiver combining all the paths of the incoming signal is called an *all-Rake* (*ARake*) receiver. Since a UWB signal has a very wide bandwidth, the number of resolvable multipath components is usually very large. Hence, an ARake receiver is not implemented in practice due to its complexity. However, it serves as a benchmark for the performance of more practical Rake receivers. A feasible implementation of multipath diversity combining can be obtained by a *selective-Rake* (*SRake*) receiver, which combines the $M$ best, out of $L$, multipath components [6]. Those $M$ best components are determined by a finger selection algorithm. For a maximal ratio combining (MRC) Rake receiver, the paths with highest signal-to-noise ratios (SNRs) are selected, which is an optimal scheme in the absence of interfering users and inter-symbol interference (ISI). For a minimum mean square error (MMSE) Rake receiver, the "conventional" finger selection algorithm is to choose the paths with highest signal-to-interference-plus-noise ratios (SINRs). This conventional scheme is not necessarily optimal since it ignores the correlation of the noise terms at different multipath components. In other words, choosing the paths with highest SINRs does not necessarily maximizes the overall SINR of the system. In [7], the optimal finger selection problem is shown to be an NP-hard problem, and two suboptimal algorithms are proposed based on an approximate objective function and constraint relaxations. In this paper, we propose a genetic algorithm (GA) based scheme, which performs finger selection by iteratively evaluating the exact objective function without the need for any constraint relaxations. Using this technique, near-optimal solutions can be obtained in many cases with a degree of complexity that is much lower than that of the optimal exhaustive search algorithm.

The remainder of this paper is organized as follows. Section II describes the transmitted and received signal models in a multiuser frequency-selective environment. The finger selection problem is formulated and the optimal algorithm is described in Section III, followed by a brief description of the conventional algorithm in Section IV. In Section V, the GA-based finger selection scheme is presented. Simulation results are presented in Section VI, and concluding remarks are made in the last section.

## II. SIGNAL MODEL

We consider a $K$-user IR-UWB system, in which the transmitted signal from user $k$ is represented by:

$$s_{\text{tx}}^{(k)}(t) = \sqrt{\frac{E_k}{N_f}} \sum_{j=-\infty}^{\infty} d_j^{(k)} b_{\lfloor j/N_f \rfloor}^{(k)} p_{\text{tx}}(t - jT_f - c_j^{(k)}T_c), \quad (1)$$

where $p_{\text{tx}}(t)$ is the transmitted UWB pulse, $E_k$ is the bit energy of user $k$, $T_f$ is the "frame" time, $N_f$ is the number of pulses representing one information symbol, and $b_{\lfloor j/N_f \rfloor}^{(k)} \in \{+1, -1\}$ is the binary information symbol transmitted by user $k$. In order to allow the channel to be shared by many users and avoid catastrophic collisions, a TH sequence $\{c_j^{(k)}\}$, where

[1]This research is supported in part by the National Science Foundation under grants ANI-03-38807, CNS-0417603, and CCR-0440443, and in part by the New Jersey Center for Wireless Telecommunications.

$c_j^{(k)} \in \{0, 1, ..., N_c - 1\}$, is assigned to each user. This TH sequence provides an additional time shift of $c_j^{(k)} T_c$ seconds to the $j$th pulse of the $k$th user where $T_c$ is the chip interval and is chosen to satisfy $T_c \leq T_f/N_c$ in order to prevent the pulses from overlapping. We assume $T_f = N_c T_c$ without loss of generality. The random polarity codes $d_j^{(k)}$ are binary random variables taking values $\pm 1$ with equal probability [8]-[10].

We assume a synchronous system and a tapped delay line channel with tap spacing $T_c$. Note that this channel model can represent any channel of the form $\sum_{l=1}^{\hat{L}} \hat{\alpha}_l^{(k)} \delta(t - \hat{\tau}_l^{(k)})$ if the channel is bandlimited to $1/T_c$ [11]. Let $\boldsymbol{\alpha}^{(k)} = [\alpha_1^{(k)} \cdots \alpha_L^{(k)}]$ represent the discrete channel for user $k$, where $L$ is assumed to be the number of multipath components for each user. Then, the received signal can be expressed as

$$r(t) = \sum_{k=1}^{K} \sqrt{\frac{E_k}{N_f}} \sum_{j=-\infty}^{\infty} \sum_{l=1}^{L} \alpha_l^{(k)} d_j^{(k)} b_{\lfloor j/N_f \rfloor}^{(k)}$$
$$\times p_{\mathrm{rx}}(t - jT_f - c_j^{(k)} T_c - (l-1)T_c) + \sigma_n n(t), \quad (2)$$

where $p_{\mathrm{rx}}(t)$ is the received unit-energy UWB pulse, and $n(t)$ is zero mean white Gaussian noise with unit spectral density.

We assume that the TH sequence is constrained to the set $\{0, 1, \ldots, N_T - 1\}$, where $N_T \leq N_c - L$, so that there is no inter-frame interference (IFI). However, the proposed algorithm is valid for scenarios with IFI as well, and this assumption is made merely to simplify the expressions throughout the paper. From the analysis in [12], the results of this paper can easily be extended to the IFI case as well.

Because of the high resolution of UWB signals, it is desirable to employ symbol-rate sampling instead of chip-rate or frame-rate sampling at the receiver. In order to enable symbol-rate sampling, the received signal is correlated with a symbol-length template signal, and the correlator output is sampled once per symbol [13]. The template signal for the $l$th path of the incoming signal is given by

$$s_{\mathrm{temp},l}^{(1)}(t) = \sum_{j=iN_f}^{(i+1)N_f - 1} d_j^{(1)} p_{\mathrm{rx}}(t - jT_f - c_j^{(1)} T_c - (l-1)T_c), \quad (3)$$

for the $i$th information symbol, where user 1 is considered as the desired user, without loss of generality. Note that the use of such template signals results in equal gain combining (EGC) of different frame components, which may not be optimal under some conditions [12]. However, it is very practical since it facilitates symbol-rate sampling. Since we consider a system that employs template signals of the form (3), i.e. EGC of frame components, it is sufficient to consider the problem of selection of the optimal paths for just one frame. Hence, we assume $N_f = 1$ without loss of generality.

Figure 1 shows the receiver structure, which uses one correlator for each multipath component. The outputs of the correlators are sampled at the symbol rate. Let $\mathcal{L} = \{l_1, \ldots, l_M\}$ denote the set of multipath components that the receiver collects. From (2) and (3), the discrete signal for the

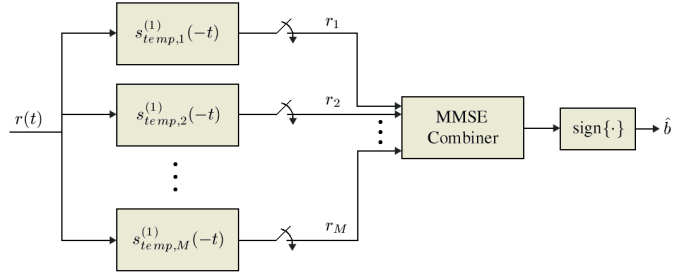

Fig. 1. The receiver structure. There are $M$ multipath components that are combined by the MMSE combiner.

$l$th path can be expressed, for the $i$th information symbol, as[2]

$$r_l = \mathbf{s}_l^T \mathbf{A} \mathbf{b}_i + n_l, \quad (4)$$

for $l = l_1, \ldots, l_M$, where $\mathbf{A} = \mathrm{diag}\{\sqrt{E_1}, \ldots, \sqrt{E_K}\}$, $\mathbf{b}_i = [b_i^{(1)} \cdots b_i^{(K)}]^T$ and $n_l \sim \mathcal{N}(0, \sigma_n^2)$. $\mathbf{s}_l$ is a $K \times 1$ vector, which can be expressed as a sum of the desired signal part (SP) and multiple-access interference (MAI) terms:

$$\mathbf{s}_l = \mathbf{s}_l^{(\mathrm{SP})} + \mathbf{s}_l^{(\mathrm{MAI})}, \quad (5)$$

where the $k$th elements can be expressed as

$$\left[\mathbf{s}_l^{(\mathrm{SP})}\right]_k = \begin{cases} \alpha_l^{(1)}, & k = 1 \\ 0, & k = 2, \ldots, K \end{cases} \quad (6)$$

and

$$\left[\mathbf{s}_l^{(\mathrm{MAI})}\right]_k = \begin{cases} 0, & k = 1 \\ d_1^{(1)} d_1^{(k)} \sum_{m=1}^{L} \alpha_m^{(k)} I_{l,m}^{(k)}, & k = 2, \ldots, K \end{cases}, \quad (7)$$

with $I_{l,m}^{(k)}$ being the indicator function that is equal to 1 if the $m$th path of user $k$ collides with the $l$th path of user 1, and 0 otherwise.

## III. OPTIMAL FINGER SELECTION

We aim to find the optimal set of multipath components, $\mathcal{L} = \{l_1, \ldots, l_M\}$, that maximizes the overall SINR of the system. In other words, we need to choose the best samples from the $L$ received samples $r_l$, $l = 1, \ldots, L$, in (4).

In order to reformulate this combinatorial problem, we first define an "assignment vector" $\mathbf{x}$, the $i$th element of which is equal to 1 if the $i$th multipath component is selected, and 0 otherwise. Since $M$ multipath components are selected by the Rake receiver, $\mathbf{x}$ satisfies $\sum_{i=1}^{L} [\mathbf{x}]_i = M$, where $[\mathbf{x}]_i$ denotes the $i$th element of $\mathbf{x}$. Also let $\mathbf{p_x}$ denote a length $M$ vector, the elements of which are the indices of the non-zero elements of $\mathbf{x}$. For example, if the second and the third multipath components are selected for a system with $L = 4$ and $M = 2$, then $\mathbf{x} = [0\ 1\ 1\ 0]$ and $\mathbf{p_x} = [2\ 3]$.

---
[2]Note that the dependence of $r_l$ on the index of the information symbol, $i$, is not shown explicitly.

From the assignment vector $\mathbf{x}$, we define an $M \times L$ "selection matrix" $\mathbf{X}$ as follows:
$$\mathbf{X} = \begin{bmatrix} \mathbf{e}_{[\mathbf{p_x}]_1} \cdots \mathbf{e}_{[\mathbf{p_x}]_M} \end{bmatrix}^T, \qquad (8)$$
where $\mathbf{e}_i$ is an $L \times 1$ unit vector having a 1 at its $i$th position and zero elements for all other entries, and $[\mathbf{p_x}]_i$ represents the $i$th element of $\mathbf{p_x}$.

Using the selection matrix $\mathbf{X}$, we can express the vector of received samples from any $M$ multipath components as
$$\mathbf{r} = \mathbf{XSAb}_i + \mathbf{Xn}, \qquad (9)$$
where $\mathbf{n}$ is the vector of thermal noise components $\mathbf{n} = [n_1 \cdots n_L]^T$, and $\mathbf{S}$ is the signature matrix given by $\mathbf{S} = [\mathbf{s}_1 \cdots \mathbf{s}_L]^T$, with $\mathbf{s}_l$ as in (5).

From (5)-(7), (9) can be expressed as
$$\mathbf{r} = b_i^{(1)}\sqrt{E_1}\mathbf{X}\boldsymbol{\alpha}^{(1)} + \mathbf{XS}^{(\mathrm{MAI})}\mathbf{Ab}_i + \mathbf{Xn}, \qquad (10)$$
where $\mathbf{S}^{(\mathrm{MAI})}$ is the MAI part of the signature matrix $\mathbf{S}$.

Then, the linear MMSE receiver can be expressed as
$$\hat{b}_i = \mathrm{sign}\{\boldsymbol{\theta}^T \mathbf{r}\}, \qquad (11)$$
where the MMSE weight vector is given by [14]
$$\boldsymbol{\theta} = \mathbf{R}^{-1}\mathbf{X}\boldsymbol{\alpha}^{(1)}, \qquad (12)$$
with $\mathbf{R}$ being the correlation matrix of the noise term:
$$\mathbf{R} = \mathbf{XS}^{(\mathrm{MAI})}\mathbf{A}^2(\mathbf{S}^{(\mathrm{MAI})})^T\mathbf{X}^T + \sigma_n^2 \mathbf{I}. \qquad (13)$$

The overall SINR of the system can be expressed as [7]
$$\mathrm{SINR}(\mathbf{X}) = \frac{E_1}{\sigma_n^2}(\boldsymbol{\alpha}^{(1)})^T \mathbf{X}^T$$
$$\left( \mathbf{I} + \frac{1}{\sigma_n^2}\mathbf{XS}^{(\mathrm{MAI})}\mathbf{A}^2(\mathbf{S}^{(\mathrm{MAI})})^T\mathbf{X}^T \right)^{-1} \mathbf{X}\boldsymbol{\alpha}^{(1)}. \qquad (14)$$

Hence, the optimal finger selection problem can be formulated as finding $\mathbf{X}$ that maximizes the SINR expression in (14), subject to the constraint that $\mathbf{X}$ has the previously defined structure. Note that the objective function to be maximized is not concave and the optimization variable $\mathbf{X}$ takes binary values, with the previously defined structure. Hence, the problem is NP-hard.

## IV. Conventional Algorithm

Instead of the solving the optimal finger selection problem, the "conventional" finger selection algorithm chooses the $M$ paths with largest individual SINRs, where the SINR for the $l$th path can be expressed as
$$\mathrm{SINR}_l = \frac{E_1(\alpha_l^{(1)})^2}{(\mathbf{s}_l^{(\mathrm{MAI})})^T \mathbf{A}^2 \mathbf{s}_l^{(\mathrm{MAI})} + \sigma_n^2}, \qquad (15)$$
for $l = 1, \ldots, L$.

This algorithm is not optimal since it ignores the correlation of the noise components of different paths, which is due to the MAI from the interfering users in the system. Therefore, it does not always maximize the overall SINR of the system given in (14).

## V. Finger Selection Using Genetic Algorithms

In this section, we propose a GA based finger selection approach, which directly uses the SINR expression in (14), and tries to achieve the optimal performance in an iterative fashion.

### A. Genetic Algorithm

The GA is an iterative technique for searching for the global optimum of an objective function [15]. The name comes from the fact that the algorithm models the natural selection and survival of the fittest [16].

The GA starts with a population of chromosomes, where each chromosome is represented by a binary string[3]. Let $N_{\mathrm{ipop}}$ denote the number of chromosomes in this population. Then, the fittest $N_{\mathrm{pop}}$ of these chromosomes are selected, according to a fitness function. After that, the fittest $N_{\mathrm{good}}$ chromosomes, which are also called the "parents", are selected and paired among themselves (*pairing* step). From each chromosome pair, two new chromosomes are generated, which is called the *mating* step. In other words, the new population consists of $N_{\mathrm{good}}$ parent chromosomes and $N_{\mathrm{good}}$ children generated from the parents by mating. After the mating step, the *mutation* stage follows, where some chromosomes (the fittest one in the population can be excluded) are chosen randomly and are slightly modified; that is, some bits in the selected binary string are flipped. After that, the pairing, mating and mutation steps are repeated until a threshold criterion is met.

The GA has been applied to a variety of problems in different areas [15]-[17]. Also, it has recently been employed in the multiuser detection problem [18]-[20]. The main characteristics of the GA algorithm is that it can get close to the optimal solution with low complexity, if the steps of the algorithm are designed appropriately.

### B. Finger Selection via the GA

In order to be able to employ the GA for the finger selection problem we need to consider how to represent the chromosomes, and how to implement the steps of the iterative optimization scheme.

A natural way to represent a chromosome is to consider the assignment vector $\mathbf{x}$ defined in Section III, which denotes the assignments of the multipath components to the $M$ fingers of the RAKE receiver. In other words, $[\mathbf{x}]_i = 1$ if the $i$th path is selected, and $[\mathbf{x}]_i = 0$ otherwise; and $\sum_{i=1}^{L}[\mathbf{x}]_i = M$.

Also, the fitness function that should be maximized can be the SINR expression given by (14). Note that, given a value of $\mathbf{x}$, $\mathrm{SINR}(\mathbf{X})$ can be uniquely evaluated. By choosing this fitness function, the fittest chromosomes of the population correspond to the assignment vectors with the largest SINR values.

Now the pairing, mating and mutation steps need to be designed for the finger selection problem:

---
[3]Although we consider only the binary GA, continuous parameter GAs are also available [15].

*1) Pairing:* The assignments to be paired among themselves are chosen according to a weighted random pairing scheme [15], where each assignment is chosen with a probability that is proportional to its SINR value. In this way, the assignments with large SINR values have a greater chance of being chosen as the parents for the new assignments.

*2) Mating:* From each assignment pair, two new pairs are generated in the following manner: Let $\mathbf{x}_1$ and $\mathbf{x}_2$ denote two finger assignments, and let $\mathbf{p}_{\mathbf{x}_1}$ and $\mathbf{p}_{\mathbf{x}_2}$ consist of the indices of the multipath components chosen as the Rake fingers. Then, the indices of the new assignments are chosen randomly from the vector $\mathbf{p} = [\mathbf{p}_{\mathbf{x}_1} \ \mathbf{p}_{\mathbf{x}_2}]$. If the new assignment is the same as $\mathbf{x}_1$ or $\mathbf{x}_2$, then the procedure is repeated for that assignment.

For example, consider a case where $L = 10$ and $M = 4$. If $\mathbf{x}_1 = [1\ 0\ 0\ 1\ 0\ 0\ 1\ 1\ 0\ 0]$ and $\mathbf{x}_2 = [0\ 1\ 0\ 1\ 0\ 1\ 0\ 0\ 1\ 0]$; that is, $\mathbf{p}_{\mathbf{x}_1} = [1\ 4\ 7\ 8]$ and $\mathbf{p}_{\mathbf{x}_2} = [2\ 4\ 6\ 9]$, then the new assignments are chosen randomly from the set $\mathbf{p} = [1\ 4\ 7\ 8\ 2\ 4\ 6\ 9]$. For example, the new assignments (children) could be $\mathbf{x}_3 = [1\ 1\ 0\ 1\ 0\ 0\ 0\ 0\ 1\ 0]$ and $\mathbf{x}_4 = [0\ 0\ 0\ 1\ 0\ 1\ 1\ 0\ 1\ 0]$ (corresponding to $\mathbf{p}_{\mathbf{x}_3} = [1\ 2\ 4\ 9]$ and $\mathbf{p}_{\mathbf{x}_4} = [4\ 6\ 7\ 9]$, respectively).

Note that by designing such a mating algorithm, we make sure that a multipath component that is selected by both parents has a larger probability of being selected by the new assignment than a multipath component that is selected by only one parent does.

*3) Mutation:* In the mutation step, an assignment, except the best one (the one with the highest SINR), is randomly selected, and one 1 and one 0 of that assignment are randomly chosen and flipped. This mutation operation can be repeated a number of times for each iteration. The number of mutations can be determined beforehand, or it might be defined as a random variable.

Now, we can summarize our GA based finger selection scheme as follows:
- Generate $N_{\text{ipop}}$ different assignments randomly.
- Select $N_{\text{pop}}$ of them with the largest SINR values.
- **Pairing:** Pair $N_{\text{good}}$ of the finger assignments according to the weighted random scheme.
- **Mating:** Generate two new assignments from each pair.
- **Mutation:** Change the finger locations of some assignments randomly except for the best assignment.
- Choose the assignment with the highest SINR if the threshold criterion is met; go to the pairing step otherwise.

In the simulations, we stop the algorithm after a certain number of iterations. In other words, the threshold criterion is that the number of iterations exceeds a given value. As the number of iterations increases, the performance of the algorithm increases, as well. The other parameters that determine the tradeoff between complexity and performance are $N_{\text{ipop}}$, $N_{\text{pop}}$, $N_{\text{good}}$, and the number of mutations at each iteration.

In terms of the computational complexity, the algorithm needs at most $N_{\text{ipop}} + N_{\text{iter}}(N_{\text{good}} + N_{\text{mut}})$ calculations of the SINR expression in (14), where $N_{\text{iter}}$ is the number of iterations, and $N_{\text{mut}}$ is the number of mutations. On the other hand, the exhaustive search for the optimal solution requires SINR calculations for $\binom{L}{M}$ different assignments.

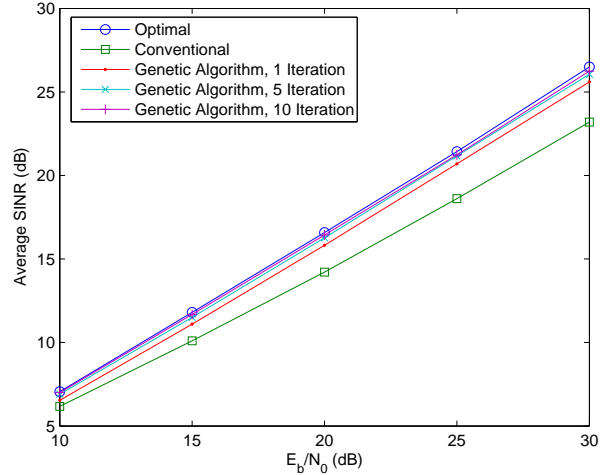

Fig. 2. Average SINR versus $E_b/N_0$ for $M = 5$ fingers, where $E_b$ is the bit energy. The channel has $L = 15$ multipath components and the taps are exponentially decaying. The IR-UWB system has $N_c = 20$ chips per frame and $N_f = 1$ frame per symbol. There are 5 equal energy users in the system and random TH and polarity codes are used.

## VI. SIMULATION RESULTS

Simulations have been performed to evaluate the performance of various finger selection algorithms for an IR-UWB system with $N_c = 20$ and $N_f = 1$. In these simulations, there are five users in the system ($K = 5$) and the users' TH and polarity codes are randomly generated. We model the channel coefficients as $\alpha_l = \text{sign}(\alpha_l)|\alpha_l|$ for $l = 1, \ldots, L$, where $\text{sign}(\alpha_l)$ is $\pm 1$ with equal probability and $|\alpha_l|$ is distributed lognormally as $\mathcal{LN}(\mu_l, \sigma^2)$. Also the energy of the taps is exponentially decaying as $\mathrm{E}\{|\alpha_l|^2\} = \Omega_0 e^{-\lambda(l-1)}$, where $\lambda$ is the decay factor and $\sum_{l=1}^{L} \mathrm{E}\{|\alpha_l|^2\} = 1$ (so $\Omega_0 = (1 - e^{-\lambda})/(1 - e^{-\lambda L})$). For the channel parameters, we choose $\lambda = 0.1$, $\sigma^2 = 0.5$ and $\mu_l$ can be calculated from $\mu_l = 0.5\left[\ln(\frac{1-e^{-\lambda}}{1-e^{-\lambda L}}) - \lambda(l-1) - 2\sigma^2\right]$, for $l = 1, \ldots, L$. We average the overall SINR of the system over different realizations of channel coefficients, TH and polarity codes of the users.

In Figure 2, we plot the average SINR of the system for different noise variances when $M = 5$ fingers are to be chosen out of $L = 15$ multipath components, and all the users have equal energy ($E_k = 1\ \forall k$). For the GA, $N_{\text{ipop}} = 32$, $N_{\text{pop}} = 16$, and $N_{\text{good}} = 8$ are used, and 8 mutations are performed at each iteration. As is observed from the figure, the GA based scheme performs considerably better than the conventional scheme, and gets very close to the optimal exhaustive search scheme after 10 iterations. The GA scheme needs to evaluate the SINR expression less than 200 times for the 10 iterations case, whereas the optimal algorithms needs 3003 evaluations. Note that the gain achieved by using the proposed algorithm over the conventional one increases as the thermal noise decreases. This is because when the thermal noise becomes less significant, the MAI becomes dominant, and the conventional technique gets worse since it ignores the correlation between the MAI noise terms when choosing the fingers.

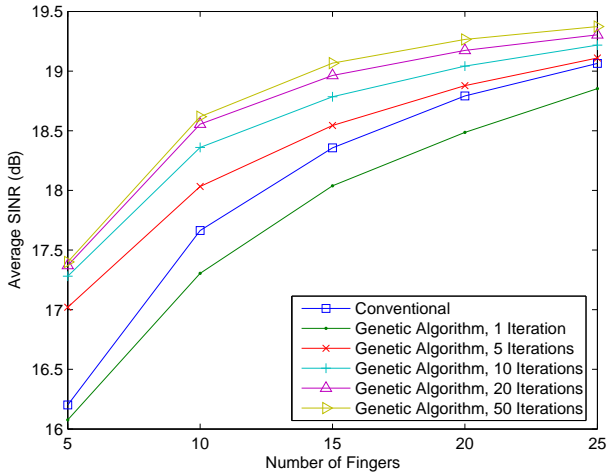

Fig. 3. Average SINR versus number of fingers $M$, for $E_b/N_0 = 20$dB, $N_c = 75$ and $L = 50$. All the other parameters are the same as those for Figure 2.

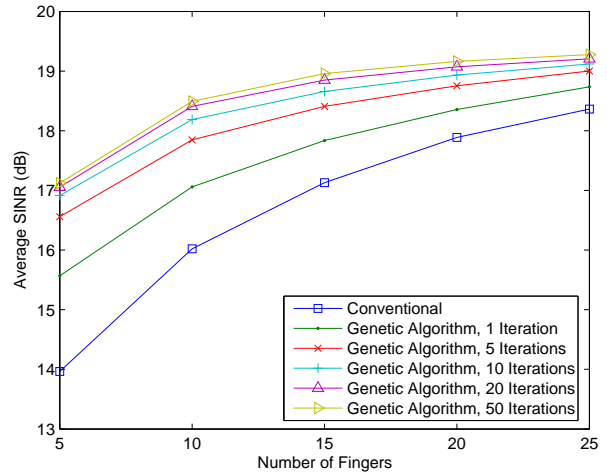

Fig. 4. Average SINR versus number of fingers $M$. There are 5 users with each interferer having 10dB more power than the desired user. All the other parameters are the same as those for Figure 3.

Next, we plot the SINR of the proposed and conventional techniques for different numbers of fingers in Figure 3, where there are 50 multipath components and $E_b/N_0 = 20$dB. The number of chips per frame, $N_c$, is set to 75, and all other parameters are kept the same as before. In this case, the optimal algorithm takes a very long time to simulate since it needs to perform exhaustive search over many different finger combinations and therefore it was not implemented. The improvement using the GA based scheme over the conventional one decreases as $M$ increases since the channel is exponentially decaying and most of the significant multipath components are already combined by both of the algorithms. The GA based scheme results in about a 1dB improvement for $M = 5$ after 10 iterations with $N_{\text{ipop}} = 128$, $N_{\text{pop}} = 64$, $N_{\text{good}} = 32$, and 32 mutations. The improvement is not significant since the MAI is not very strong in this case.

Finally, we consider an MAI-limited scenario, in which there are 5 users with $E_1 = 1$ and $E_k = 10 \ \forall k \neq 1$, and all the parameters are as in the previous case. Then, as shown in Figure 4, the improvement by using the proposed algorithm increases significantly. The main reason for this is that the GA based scheme considers the correlations caused by MAI whereas the conventional scheme simply ignores it.

## VII. CONCLUDING REMARKS

Since UWB systems have a large numbers of multipath components, only a subset of those components can be used due to complexity constraints. Therefore, the selection of the optimal subset of multipath components is important for the performance of the receiver. The optimal solution to this finger selection problem requires exhaustive search which would become prohibitive for UWB systems. Therefore, we have proposed a GA based iterative finger selection scheme, which depends on the direct evaluation of the objective function. In each iteration, the set of possible finger assignments is updated in search of the best assignment according to the proposed GA stages.